\documentclass[12pt,twoside]{article}
\usepackage{fleqn,espcrc1}
\usepackage{graphicx}
\usepackage{epsfig}
\newcommand{\be}{\begin{eqnarray}}

\newcommand{\ee}{\end{eqnarray}}
\newcommand{\ben}{\begin{eqnarray*}}
\newcommand{\een}{\end{eqnarray*}}
\newcommand{\stackeven}[2]{{{}_{\displaystyle{#1}}\atop\displaystyle{#2}}}
\newcommand{\lsim}{\stackeven{<}{\sim}}

\newcommand{\as}{\alpha_s}
\def\eq#1{{Eq.~(\ref{#1})}}
\def\fig#1{{Fig.~\ref{#1}}}

\newcommand{\AmS}{{\protect\the\textfont2
  A\kern-.1667em\lower.5ex\hbox{M}\kern-.125emS}}

\title{
Classical Initial Conditions for Nucleus-Nucleus Collisions}

\author{Yuri V.\ Kovchegov \address{Department of Physics, 
University of Washington, Box 351560 \\ Seattle, WA 98195-1560, USA}%
\thanks{This work has been supported in part by the U.S. Department of
Energy under Grant No. DE-FG03-97ER41014.}}

\begin{document}

\maketitle

\begin{abstract}
We present the results of an analytical calculation of the
distribution of gluons in the state immediately following a heavy ion
collision in the quasi--classical limit of QCD given by
McLerran--Venugopalan model. We show that the typical transverse
energy $E_T$ of the produced gluons is of the order of the saturation
scale of the nuclei $Q_s$, as predicted by Mueller. We demonstrate
that due to multiple rescatterings the obtained gluon distribution
remains finite (up to logarithms of $k_\perp$) in the soft transverse
momentum limit of $k_\perp \, \ll \, Q_s$. We compare the gluon
spectrum in the single nuclear wave function before the collision to
the spectrum of the produced gluons. The total number of produced
gluons turns out to be proportional to the total number of gluons
inside the nuclear wave function before the collision with the
proportionality coefficient $c \, \approx \, 2 \, \ln 2$.\\
\end{abstract}
\renewcommand{\thefootnote}{\arabic{footnote}}
\setcounter{footnote}{0}


The results presented here are based on reference \cite{me}. Let us
consider a collision of two ultrarelativistic nuclei characterized by
the saturation scales $Q_{s1}^2$ and $Q_{s2}^2$. We are going to write
down an analytical expression for the classical distribution of
produced gluons which resums all multiple rescatterings of the gluons
with the nucleons in the colliding nuclei. Formally this corresponds
to resumming powers of both $Q_{s1}^2 / k_\perp^2$ and $Q_{s2}^2 /
k_\perp^2$ \cite{mv,meM}. The problem is similar to the problem of
finding the classical gluon field of two colliding nuclei formulated
in \cite{kmw}, with the exception that here we are interested in the
gluon production cross section.

The classical field of two nuclei in the forward light cone has been
found in the usual perturbation theory to order $g^3$ in
\cite{kmw}. The answer for the gluon multiplicity distribution is the 
first (lowest order) term of the expansion in powers of $Q_{s1,2}^2 /
k_\perp^2$ (see \eq{uv} below) and corresponds in this sense to
proton-proton scattering.

A gluon production cross section for a slightly more complicated case
of proton-nucleus interactions was derived in \cite{meM}. Formally
that cross section includes all powers of $Q_{s1}^2 / k_\perp^2$
keeping only the leading power of $Q_{s2}^2 / k_\perp^2$. The
calculations in \cite{meM} were done in covariant gauge, where
multiple {\it final} state rescatterings play crucial role in gluon
production.

\begin{figure}
\begin{center}
\epsfxsize=15cm
\leavevmode
\hbox{ \epsffile{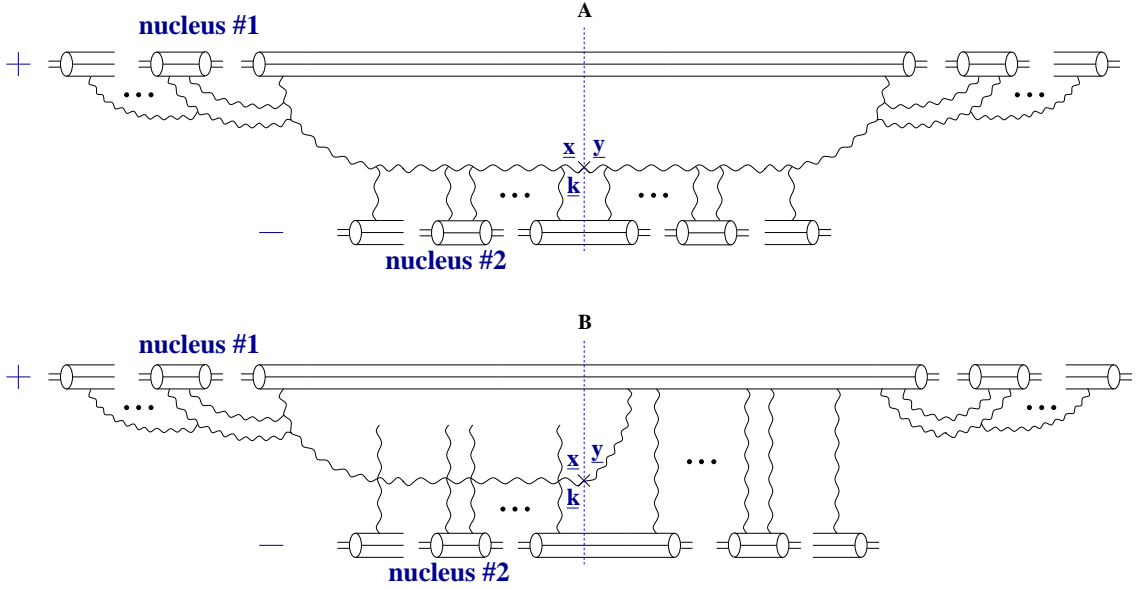}}
\end{center}
\caption{Diagrams contributing to the gluon production in nucleus-nucleus 
collisions in the $A_+ = 0$ light cone gauge.}
\label{aafig}
\end{figure}

In \cite{me} the same pA process was considered in the framework of
light cone perturbation theory in light cone gauge. Knowing the answer
for the gluon production cross section from \cite{meM} allowed us to
determine diagrams which reproduce this answer in the light cone
gauge. We demonstrated that an entirely different set of {\it initial}
state interactions is important there. Generalizing the pA results to
the AA case we end up with the diagrams shown in
Fig. \ref{aafig}. Summation of the graphs of Fig. \ref{aafig} yields
us with the following multiplicity distribution of the produced gluons
\cite{me}
\ben
\frac{d N^{AA}}{d^2 k \, d^2 b \, dy} \ = \ \frac{2 \, C_F}{\as \pi^2} \, 
\left\{  -  \int \frac{d^2 z}{(2 \pi)^2} \,  e^{i {\underline k} \cdot 
{\underline z}} \, \frac{1}{{\underline z}^2} \, \left(1 - e^{ -
{\underline z}^2 \, Q_{s1}^2 / 4 } \right) \, \left(1 - e^{ -
{\underline z}^2 \, Q_{s2}^2 / 4 } \right) + \right.
\een
\ben
+ \left. \int \frac{d^2 x \ d^2
y}{(2 \pi)^3} \, e^{i {\underline k} \cdot ({\underline x} -
{\underline y})} \,  \frac{{\underline x}}{{\underline x}^2}
\cdot \frac{{\underline y}}{{\underline y}^2} \,  \left[ \frac{1}{{\underline x}^2 
\, \ln \frac{1}{|{\underline x}| \mu}} \, \left(1 - e^{ - {\underline x}^2 
\, Q_{s1}^2 / 4 } \right) \, \left(1 - e^{ - {\underline x}^2
\, Q_{s2}^2 / 4 } \right) + \right. \right.
\een
\be\label{aasol}
+ \left. \left.  \frac{1}{{\underline y}^2 
\, \ln \frac{1}{|{\underline y}| \mu}} \, \left(1 - e^{ - {\underline y}^2 
\, Q_{s1}^2 / 4 } \right) \, \left(1 - e^{ - {\underline y}^2
\, Q_{s2}^2 / 4 } \right) \right] \right\}.
\ee
\eq{aasol} is our main result. It provides us with the number of gluons 
produced in a heavy ion collision per unit transverse momentum phase
space, per unit rapidity interval at the given impact parameter $b$
yielding the initial conditions for possible thermalization of the
gluonic system at later times.

\begin{figure}
\begin{center}
\epsfxsize=9cm
\leavevmode
\hbox{ \epsffile{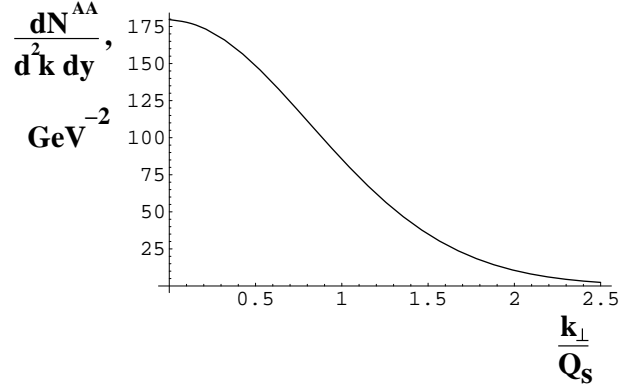}}
\end{center}
\caption{Distribution of the produced gluons given by \eq{dist2} for a 
central AA collision as a function of $k/Q_s$ with the transverse area
of the nuclei being $S_\perp = 50 \, \mbox{fm}^2$. The approximation
of \eq{dist2} is valid only for $k_\perp \, \lsim \, Q_s$.}
\label{dist}
\end{figure}

To explore the properties of the distribution (\ref{aasol}) we first
perform the transverse coordinate integration in the logarithmic
approximation (see \cite{me}), which would transform it into the
following simplified form
\be\label{dist2}
\frac{d N^{AA}}{d^2 k \, d^2 b \, dy} \ = \ \frac{C_F}{\as 2 \pi^3} \, 
\frac{1}{{\underline k}^2} \, \left[ (Q_{s1}^2 + Q_{s2}^2) \, e^{- 
\frac{{\underline k}^2}{Q_{s1}^2 + Q_{s2}^2}} - Q_{s1}^2 \, e^{- 
\frac{{\underline k}^2}{Q_{s1}^2}} - Q_{s2}^2 \, e^{- 
\frac{{\underline k}^2}{Q_{s2}^2}}\right].
\ee
The distribution of \eq{dist2} is plotted in \fig{dist} as a function
of $k/Q_s$ for the case of two identical cylindrical nuclei with
$Q_{s1} = Q_{s2} = Q_s$ and with the cross sectional area $S_\perp \,
= \, \pi \, R^2 \, \approx \, 50 \, \mbox{fm}^2 \, \approx \, 1250 \,
\mbox{GeV}^{-2}$. Note that \eq{dist2} is valid only in the not 
very large transverse momentum region $k_\perp \lsim Q_s$ \cite{me}.

As one can see, the distribution in \fig{dist} remains finite as
$k_\perp/Q_s \rightarrow 0$. If one takes the $k_\perp \, \ll \, Q_s$
limit of \eq{dist2} then the distribution goes to a constant
\be\label{ir}
\frac{d N^{AA}}{d^2 k \, d^2 b \, dy} \ \rightarrow \ \ \frac{C_F}{\as 2 \pi^3} 
\hspace*{1cm} \mbox{as} \hspace*{1cm} \frac{k_\perp}{Q_s} \, \ll \, 1.
\ee 
This means that the exact expression of \eq{aasol} may only have
logarithmic divergences in the infrared limit. Therefore, finiteness
of \eq{dist2} at small transverse momenta demonstrates that multiple
rescatterings are the reason the hadronic and nuclear single particle
inclusive production cross sections remain finite in the soft momentum
region.

Since the distribution of \eq{aasol} includes the lowest order in
$\as$ diagrams in it, one readily derives that in the $k_\perp/Q_s
\rightarrow \infty$ limit the distribution falls off as $1/k_\perp^4$
\cite{kmw}
\be\label{uv}
\frac{d N^{AA}}{d^2 k \, d^2 b \, dy} \ \sim \ \frac{Q_{s1}^2 \, Q_{s2}^2}
{\as \, {\underline k}^4} \hspace*{1cm} \mbox{as} \hspace*{1cm}
\frac{k_\perp}{Q_s} \rightarrow \infty .
\ee
The average transverse energy per produced gluon given by \eq{dist2}
for the case of two identical nuclei with $Q_{s1} = Q_{s2} = Q_{s}$ is
\be
 < E_T > \, = \, \frac{\sqrt{\pi} (\sqrt{2} - 1)}{\ln 2} \, Q_s
\, \approx \, 1.06 \, Q_s .
\ee
That is, the typical transverse energy of the produced gluons is of
the order of the saturation scale, as was previously conjectured by
Mueller in \cite{Mueller2}. The saturation scale for a large nucleus
scales as $Q_s^2 \, \sim \, A^{1/3}$ with atomic number justifying the
use of the small coupling expansion for large nuclei
\cite{mv,Mueller2}.

\begin{figure}
\begin{center}
\epsfxsize=9cm
\leavevmode
\hbox{ \epsffile{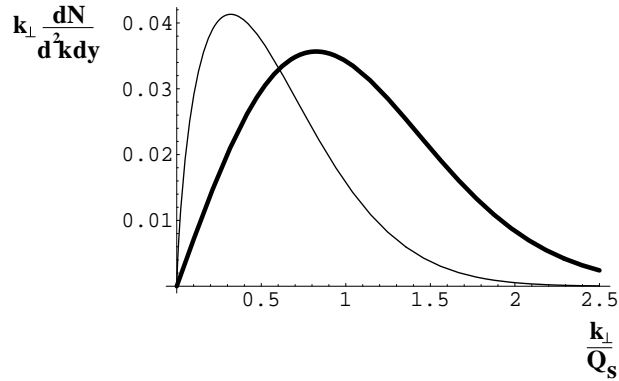}}
\end{center}
\caption{The gluon distribution as a 
function of $k_\perp/Q_s$ in one of the nuclei before the collision
(thin line) and the gluon distribution after the collision (thick
line).}
\label{ba}
\end{figure}

Fig. \ref{ba} compares the transverse momentum distribution of gluons
in the single nuclear wave function before the collision (thin line)
with the distribution of produced gluons after the collision (thick
line). One can see that the typical transverse momentum of gluons gets
broadened due to multiple rescatterings in the collision. The total
number of gluons produced in the collision can be found by integrating
\eq{dist2} over $k_\perp$. The result yields for two identical nuclei
\be\label{tot}
\frac{d N^{AA}}{d^2 b \, dy} \ = \ \frac{C_F \, Q_s^2 \, \ln 2}{\as \pi^2}.
\ee
The total number of produced gluons turns out to be proportional to
the total number of gluons in the wave function of a single nucleus
before the collision \cite {me} with the proportionality
coefficient 
\be
c \ = \ 2 \, \ln 2 \ \approx \ 1.39 .
\ee
The obtained value for the ``gluon liberation'' coefficient $c = 2 \ln
2$ is close to one, as was originally suggested by Mueller
\cite{Mueller2}.

\end{document}